\documentclass[twocolumn,floatfix,preprintnumbers,citeautoscript,newabstract,nofootinbib]{revtex4} 
\usepackage{amsmath}
\usepackage{amssymb} 
\usepackage{bbm} 
\usepackage[small]{caption2} 
\usepackage{fleqn} 
\usepackage{graphicx} 
\usepackage{mathrsfs}   
\usepackage[small,loose]{subfigure}  
\usepackage{color}

\newcommand{\eVdist}{\kern-0.06em}

\newcommand{\gev}{\:\text{Ge\eVdist V}}

\newcommand{\SO}[1]{\ensuremath{\mathrm{SO}(#1)}}
\newcommand{\SU}[1]{\ensuremath{\mathrm{SU}(#1)}}
\newcommand{\U}[1]{\ensuremath{\mathrm{U}(#1)}}
\newcommand{\Z}[1]{\ensuremath{\mathbbm{Z}_{#1}}} 

\newcommand{\qhu}{{\ensuremath{q_{H}}}}
\newcommand{\qhd}{{\ensuremath{q_{\Bar{H}}}}}

\allowdisplaybreaks[1]

\begin{document}

\preprint{TUM-HEP 770/10; LMU-ASC 64/10; OHSTPY-HEP-T-10-003; CERN-PH-TH/2010-193; OUTP-10-24P}
\title{\Large\bf
A unique $\boldsymbol{\mathbbm{Z}_4^R}$ symmetry for the MSSM
}

\vspace{1cm}

\author{\textbf{
Hyun Min Lee$^a$,
Stuart Raby$^b$,
Michael Ratz$^c$,\\
Graham G.~Ross$^{a,d}$,
Roland Schieren$^c$,
Kai Schmidt-Hoberg$^c$,\\
Patrick K.~S.~Vaudrevange$^e$
}
\\[5mm]
\textit{$^a$\small
~Theory Group, CERN, 1211 Geneva 23, Switzerland
}
\\[2mm]
\textit{$^b$\small
~Department of Physics, The Ohio State University,\\
191 W.\ Woodruff Ave., Columbus, OH 43210, USA
}
\\[2mm]
\textit{$^c$\small
~Physik-Department T30, Technische Universit\"at M\"unchen, \\
James-Franck-Stra\ss e, 85748 Garching, Germany
}
\\[2mm]
\textit{$^d$\small
~Department of Physics, Theoretical Physics, University of Oxford, \\
1 Keble Road, Oxford OX 1 3NP, U.K.
}
\\[2mm]
\textit{$^e$\small
~Arnold Sommerfeld Center for Theoretical Physics,\\
Ludwig-Maximilians-Universit\"at M\"unchen, 80333 M\"unchen, Germany
}}

\begin{abstract}
We consider the possible anomaly free Abelian discrete symmetries of the MSSM
that forbid the $\mu$-term at perturbative order. Allowing for anomaly
cancellation via the Green-Schwarz mechanism we identify discrete $R$-symmetries
as the only possibility and prove that there is a unique $\Z{4}^{R}$ symmetry
that commutes with \SO{10}.  We argue that non-perturbative effects will
generate a $\mu$-term of electroweak order thus solving the $\mu$-problem. The
non-perturbative effects break the $\Z{4}^{R}$ symmetry leaving an exact $\Z{2}$
matter parity. As a result dimension four baryon- and lepton-number violating
operators are absent while, at the non-perturbative level, dimension five baryon-
and lepton-number violating operators get induced but are highly suppressed so
that the nucleon decay rate is well within present bounds.
\end{abstract}

\maketitle

\section{Introduction}

 Supersymmetric extensions of the Standard Model (SM) are very popular as they
can solve the hierarchy problem, stabilizing the electroweak scale against a
high scale associated with new physics. The simplest such extension, the MSSM,
assumes the minimal number of new particle states. However there are several
problems that immediately arise in its construction associated with terms in the
Lagrangian that are allowed by supersymmetry and by the gauge symmetry of the
SM. At dimension three there is a Higgs mass term with coefficient $\mu$ (as
well as $R$ parity violating terms $\mu_i H L_i$), that, if unsuppressed,
reintroduces the hierarchy problem -- the so-called $\mu$-problem.  At dimension
four and five there are baryon- and lepton-number violating terms that must be
strongly suppressed to prevent unacceptably fast nucleon decay. 

 Discrete symmetries play an important role in controlling these terms. It is
normally assumed that the MSSM should also be invariant under a
$\mathbbm{Z}_{2}$ matter-parity
\cite{Farrar:1978xj,Dimopoulos:1981zb,Dimopoulos:1981dw}, that distinguishes
between lepton and down-type Higgs doublets. Such discrete symmetries may be
constrained by the requirement of anomaly freedom
\cite{Krauss:1988zc,Ibanez:1991hv,Banks:1991xj}. This is certainly the case if they come from
a spontaneously broken gauge symmetry. It is also the case if they come from a
string theory;  for instance, in orbifolds they can
arise as discrete remnants of the Lorentz group
in the compact space. 
It has also been argued that non-gauge discrete symmetries are
violated by gravitational effects rendering them ineffective~\cite{Krauss:1988zc}.

 In this paper we revisit the question what is the underlying symmetry of the
MSSM, focusing on anomaly free discrete symmetries to avoid the appearance of
new light gauge degrees of freedom. The symmetry should be capable of ensuring
that the nucleon is stable, at least within present bounds. Early attempts,
assuming the MSSM spectrum, classified low-order anomaly free Abelian discrete
symmetries, including $R$-symmetries, that can stabilize the nucleon. Matter
parity is anomaly free but allows dimension five nucleon decay
operators~\cite{Ibanez:1991pr}. A $\mathbbm{Z}_{3}$ ``baryon-triality'' is
anomaly free that forbids both dimension four baryon number violation operators
and dimension five nucleon decay operators but allows lepton number violating
dimension four operators. The combination of these two gives a $\mathbbm{Z}_{6}$
``proton hexality'' symmetry \cite{Dreiner:2005rd} that forbids all dimension
four baryon and lepton number violating operators and dimension five nucleon
decay operators. However these symmetries allow the problematic $\mu$ term;
indeed it was a requirement that it should be allowed.

 Here we adopt a different philosophy and look for anomaly free discrete
symmetries that forbid both the $\mu$ term and all dimension four and five
baryon and lepton number violating operators. Of course an electroweak scale
$\mu$ term is needed but we argue that it should arise through spontaneous
non-perturbative breaking of the discrete symmetry thus solving the
$\mu$-problem. We also require that, unlike baryon-triality or proton hexality
(cf.\ \cite{Forste:2010pf}), the discrete symmetry should 
commute with a simple Grand Unified group such as \SU{5} or \SO{10}, thus
readily preserving the attractive features of Grand or string unification. In
addition we require that all fermion masses, including neutrino masses, should
be allowed by the symmetry.
 
 Remarkably we find that in the \SO{10} case there is a unique solution, a
$\mathbbm{Z}_{4}^{R}$ $R$-symmetry. The symmetry may be broken non-perturbatively
generating a $\mu$ term of the correct order plus dimension five baryon and
lepton number violating operators that are sufficiently suppressed to be
consistent with bounds on nucleon decay.
 A $\mathbbm{Z}_{2}$ symmetry is left
unbroken, equivalent to matter parity, that forbids the generation of dimension
four baryon and lepton number violating terms and ensures the LSP is stable. The
spontaneous breaking of the discrete symmetry leads to a potential domain wall
problem but we show this is avoided provided a relatively mild constraint on the
reheat temperature after inflation is satisfied.  
 
 Anomaly cancellation proceeds via the Green-Schwarz (GS) mechanism
\cite{Ibanez:1991pr,Ibanez:1992ji}. In recent years it has become clear that
anomaly free discrete symmetries of this type need not originate from the
so-called `anomalous \U1' but can have a string origin
\cite{Dine:2004dk,Araki:2008ek} although, at least in heterotic orbifolds, there
is a tight relation between these symmetries and the anomalous
\U1~\cite{Araki:2008ek}. The GS anomaly cancellation requires that the
anomaly coefficients be universal. Because these symmetries arise from string
compactifications, the effects that violate them are well under control, and
they can be viewed as approximate symmetries.

 The paper is organized as follows. We first show that only discrete $R$
symmetries can satisfy the constraints discussed above. Then we prove that, for
the case the symmetry commutes with \SO{10}, there is a unique discrete $\Z4^R$
symmetry which allows for the usual Yukawa couplings and the Weinberg operator
generating Majorana masses for the neutrinos; anomaly cancellation proceeds via
the Green-Schwarz mechanism.  We then consider the phenomenological implications
of the model, including the cosmological implications. Finally we comment on the
possible origin of this $\Z4^R$ symmetry and briefly discuss an explicit
string-derived model with the exact MSSM spectrum below the string scale with
this symmetry.

\section{A unique $\boldsymbol{\Z4^R}$ symmetry for the MSSM}

In this section we prove that with the minimal field content of the MSSM there
is a unique discrete symmetry with the following features:
\begin{itemize}
\item[(i)] Anomaly cancellation (allowing for a GS term).
\item[(ii)] Consistency with \SO{10}.
\item[(iii)] No $\mu$ term at the perturbative level.
\item[(iv)] Quark, charged lepton and neutrino masses allowed.
\end{itemize}

 Consider a  $\Z{N}$ symmetry under which the matter superfields have charge
$q^{(f)}$. For the case $\Z{N}$ is an $R$-symmetry the fermion components have
charge $q^{(f)}-1$ under the convention that the superpotential $\mathscr{W}$
has $\mathbbm{Z}_N^R$ charge~2. In general, the anomaly coefficients are given by
\begin{align}
 A_{G-G-\Z{N}} 
 &=~
 \sum_{\boldsymbol{r}^{(f)}} \ell(\boldsymbol{r}^{(f)})\,(q^{(f)}-R)
 	+\ell(\text{adj}) \cdot R\;, 
\end{align}
where $G$ is the gauge group, $q^{(f)}$ denotes the chiral
superfield \Z{N} charge and $\ell(\text{adj})$ is the Dynkin index of the
adjoint representation of $G$. For an $R$-symmetry $R=1$, otherwise $R=0$. The
sum runs over the irreducible representations  $\boldsymbol{r}^{(f)}$ of $G$ of
the chiral fields with Dynkin index
$\ell(\boldsymbol{r}^{(f)})$.  Our conventions are such that
$\ell(\boldsymbol{M})=1/2$ for \SU{M} and $\ell(\boldsymbol{M})=1$ for \SO{M}.
The condition of anomaly cancellation corresponds to 
\begin{equation}
 A_{G-G-\Z{N}}~=~ \rho \mod \eta \;,
\end{equation}
where  
\begin{equation}\label{eq:eta}
 \eta~:=~
 \left\{\begin{array}{ll}
   N & \text{for $N$ odd}\;,\\
   N/2 & \text{for $N$ even}\;.
\end{array}\right. 
\end{equation}
In the absence of a GS term the constant $\rho=0$. Allowing for a GS term the
anomaly cancellation condition is relaxed to the condition
$A_{G-G-\Z{N}}=\rho$ for the (suitably normalised) gauge factors of the
SM. 

We now impose the condition that \Z{N} should commute with \SU{5} and assign
discrete charges $q_{{\boldsymbol{10}}_i}$ and 
$q_{{\boldsymbol{\overline{5}}}_i}$ to the ${\boldsymbol{10}}$
and
${\boldsymbol{\overline{5}}}$
representations making up family
$i$. The coefficients for the mixed $\SU{3}_C$ and $\SU{2}_\text{L}$ anomalies
are
\begin{align}
\label{eq:A-GUT-bosonic1}
 A_{\SU3-\SU3-\Z{N}} & = ~
 \frac{1}{2}\sum_{i}\left[3 \cdot q_{{\boldsymbol{10}}_i}+q_{{\boldsymbol{\overline{5}}_i}}-4R\right ] +3R\;, \\
 \label{eq:A-GUT-bosonic2}
 A_{\SU2-\SU2-\Z{N}}
 & =~ 
 \frac{1}{2}\sum_{i}\left[3 \cdot q_{{\boldsymbol{10}}_i}+q_{{\boldsymbol{\overline{5}}_i}}-4R\right ]+2R \nonumber
 \\& \quad +\frac{1}{2}\left(\qhu+\qhd-2R \right)\;,
\end{align}
where $\qhu$ and $\qhd$ denote the \Z{N} charges of the up-type and down-type Higgs doublets, $H$ and
$\Bar{H}$ respectively.

 Allowing for the GS term, anomaly cancellation (universality) requires
(cf.\ \cite{Hamaguchi:2003za})
\begin{align}
\left(\qhu+\qhd\right) & =~ 4R \mod 2\eta \;. 
\label{hc}
\end{align}
This should be compared to the condition that a Higgs mass term is allowed which
is
\begin{align}
\left(\qhu+\qhd\right) & =~ 2R \mod N \;. 
\label{mut}
\end{align} 

 We immediately see that for a non-$R$ symmetry ($R=0$) the requirement that 
\Z{N} commute with \SU{5} means that the Higgs mass term in the superpotential
is allowed  and the $\mu$-problem remains (cf.\ the similar discussion in
\cite{Hall:2002up}). 

 The situation is different for the case of an $R$ symmetry ($R=1$) and it is not
difficult to demonstrate that there are solutions to eq.~\eqref{hc} that do not
satisfy eq.~\eqref{mut} and thus solve the $\mu$-problem. In fact one can show
that any solution that forbids the dimension five nucleon decay operators also
forbids the $\mu$-term and moreover that $N$ should be a divisor of 24. We will
discuss the general case elsewhere~\cite{LRRRSSV2} but here we show that in  the
special case that \Z{N} commutes with \SO{10} there is a unique solution to the
anomaly cancellation equations and that it does solve the $\mu$-problem. In this
case $q_{{\boldsymbol{10}}_i}=q_{{\boldsymbol{\overline{5}}_i}}=q$ where, to
allow for interfamily mixing, the charges must be family independent. 

 To generate masses for the quarks, charged leptons and neutrinos  we require
that the $u$- and $d$-type Yukawa couplings and the Weinberg
operator $(L H)^{2}$ be allowed.\footnote{Allowing the Weinberg operator
is equivalent to adding right-handed neutrinos $\nu^c$ with charge 1. Then
Yukawa couplings of $\nu^c$ and Majorana mass terms are also permitted.} This
yields the following conditions between the $R$-charges:
\begin{eqnarray}
 2q+\qhu  & = & 2 \mod N\;,\\
 2q+\qhd  & = & 2 \mod N\;,\\
 2q+2\qhu & = & 2 \mod N
\end{eqnarray}
with solution
\begin{eqnarray}
 \qhu & = & \qhd~=~0 \mod N\;,\\
 2q & = & 2 \mod N\;.
\end{eqnarray}
Inserting this in eq.~\eqref{hc} shows that $N$ can take two values only, $N=2$
or $N=4$. The $\Z2^R$ symmetry does not forbid the $\mu$ term and indeed there
are no meaningful discrete $R$ symmetries of order 2 (cf.\ e.g.\
\cite{Dine:2009swa} for a recent discussion).  We are therefore left with the
unique possibility of a $\mathbbm{Z}_4^R$ symmetry  with $\Z4^R$ charges as
given in table~\ref{tab:Z4R}.
This symmetry has been considered before in \cite{Babu:2003qh} using the
Giudice-Masiero mechanism \cite{Giudice:1988yz} to generate the $\mu$ term.
Another version of this symmetry with $\rho=0$ and extra matter has been discussed in
\cite{Kurosawa:2001iq}.

\begin{table}[!h!]
\begin{center}\begin{tabular}{| ccccccc |}
\hline
$Q$ & $U^c$ & $E^c$ & $D^c$ & $L$ & $H$ & $\Bar{H}$ \\
1 & 1 & 1 & 1 & 1 & 0 & 0 \\
\hline
\end{tabular}\end{center}
\caption{$\Z{4}^R$ charge assignment for the MSSM superfields.
}
\label{tab:Z4R}
\end{table}

The last step is to check the remaining anomaly cancellation conditions. 
Mixed
$\U1$-$\U1$-$\Z{N}$ anomalies are often ignored as they do not give meaningful
constraints unless one knows the normalization of the
charges~\cite{Ibanez:1992ji,Dreiner:2005rd}.  However, in the case of
hypercharge $Y$ the normalization is fixed by the underlying GUT. The
resulting $\U1_Y$-$\U1_Y$-$\Z{N}^R$ anomaly condition is
\begin{eqnarray}
\lefteqn{ A_{\U1_Y\text{-}\U1_Y\text{-}\Z{N}^R}
 ~=~\frac{3}{5}\cdot
 \bigg\{2\cdot\left(\tfrac{1}{2}\right)^2\left[\qhu+\qhd-2\right]}\nonumber\\
 & & {}+3(q_{\boldsymbol{\overline{5}}}-1)\left[2\left(\tfrac{1}{2}\right)^{2}
 	+3\left(\tfrac{1}{3}\right)^{2}\right]\nonumber\\
 & & {}+3(q_{\boldsymbol{10}}-1)\left[6\left(\tfrac{1}{6}\right)^{2}
 +3\left(\tfrac{2}{3}\right)^{2}+(1)^{2}\right]\bigg\}\nonumber\\
 & &~=~
 \rho \mod \eta\;.\label{eq:Z4r-U1Y-U1Y}
\end{eqnarray} 
Note that this anomaly coefficient is not invariant under shifting some discrete
charges by multiples of $N$. That is, there are equivalent $\Z{N}^R$ charge
assignments, leading to different anomaly coefficients. We find that the true
anomaly constraint is that there has to exist a charge assignment under which
the conditions of anomaly freedom or anomaly universality are satisfied.
For the case of  $\Z4^R$ $\rho=1$ and $\eta=2$.
In this case, one may check that the anomaly cancellation
condition, eq.~\eqref{eq:Z4r-U1Y-U1Y}, is satisfied for the choice that both
Higgs superfields have $R$ charge $-4$ rather than 0.

The $\mathbbm{Z}_N^3$ anomaly does not yield model independent
constraints~\cite{Banks:1991xj,Araki:2008ek}; a more detailed discussion will be
presented in~\cite{LRRRSSV2}. However  there is one further anomaly cancellation
condition of interest, namely the grav-grav-$\Z{N}^R$ graviton anomaly. It too
is often ignored as it can always be satisfied by adding SM singlet fields but
it is still of some interest because the existence of additional \emph{light}
singlet states is potentially of phenomenological interest.  We have
\begin{align}
 A_{\text{grav-grav-}\Z{N}^{R}}&=~-21+8+3+1\nonumber\\& 
 +3\, \{10\cdot(q_{\boldsymbol{10}}-1)+5\, (q_{\boldsymbol{\overline{5}}}-1)\}
\nonumber \\
 &{} +2\, (\qhu+\qhd-2)~=~24\rho\mod \eta \;.
\end{align}
For the case of  $\Z4^R$ this constraint is $-9-4=24 \mod 2$ which is not
satisfied. Thus there must be additional SM singlet state(s) with $\Z{4}^R$
charges $s_{i}$ such that $\sum_{i}(s_{i}-1)$ is odd. We will discuss what mass
they may acquire shortly. 

\section{$\boldsymbol{\Z4^R}$ phenomenology}
\label{sec:simpleZ4R}

Clearly, the charge assignment given in table~\ref{tab:Z4R} is consistent with
Grand Unification for matter. In \SO{10} language it corresponds to giving the
$\boldsymbol{16}$-plet a $\Z4^R$ charge 1, such that the matter fermions
transform trivially, and $\Z4^R$ charge $-4\,\widehat{=}\,0$ to the Higgs fields
contained in the $\boldsymbol{10}$-plet. Notice that successful doublet-triplet
splitting is required for these anomalies to be universal, \emph{i.e.}\ the
$\Z4^R$ does not commute with \SO{10} in the Higgs sector. This is the usual
doublet-triplet splitting problem that is most elegantly solved in string
unification via Wilson line breaking.

The structure of the renormalisable terms of the $\Z4^R$ model is identical to
that of the usual MSSM with matter parity with the exception that the $\mu$ term
is absent. The terms of dimension five differ in that the baryon- and
lepton-number violating terms $QQQL$ and $U^c U^c D^c E^c$ are absent.
However the dimension five Weinberg operator $(LH)^{2}$ is allowed and this
generates Majorana masses for the neutrinos. In an underlying Grand Unified
theory we expect these terms to be generated by the usual see-saw mechanism.

 Of course the critical question is, how the $\mu$-term is generated. There are
two ways this can happen, either by a $D$-term of the form
$X^{{\dagger}}H\Bar{H}$ \cite{Giudice:1988yz} or via an $F$-term of the form
$YH\Bar{H}$ \cite{Kim:1983dt} where $X$, $Y$ may be a single field or a
composite operator (cf.\ also the discussion in~\cite{Choi:1996fr}). The
$\Z4^{R}$-charge of $X$ and $Y$ must be $0$ and $2$ respectively and a
$\mu$-term is generated if the $F$-term of $X$ or the $A$ term of $Y$ acquires a
vacuum expectation value (VEV).  Both cases break the $\Z4^{R}$ symmetry leaving
a $\Z2$ symmetry unbroken that, together with invariance of the Lagrangian under
a change of sign of the fermion fields, is equivalent to the usual $R$-parity of
the MSSM.  Such a breaking is necessary to allow for gaugino masses. In fact we
expect such breaking of the symmetry to occur through non-perturbative effects
since the $\Z4^R$ is anomalous in the absence of a GS term.

 As discussed in the next section, a plausible origin of such non-perturbative
terms is through a hidden sector that dynamically  generates a vacuum
expectation value for the superpotential via a gaugino condensate. This
corresponds to identifying $Y$ with $\mathscr{W}$, the latter being the order
parameter of $\Z4^{R}$ breaking. In this case the $\mu$ term is of
$\mathcal{O}(\langle\mathscr{W}\rangle/M^{2}$) where $M$ is the messenger field
mass. For the case of gravity mediation (SUGRA) $M$ is the Planck mass and
$\mathscr{W}$ is also the order parameter for SUGRA;
$\langle\mathscr{W}\rangle/M_\mathrm{P}^{2}=m_{3/2}$ is the gravitino mass. A specific
realization in string theory will be briefly discussed in the next section and
in more detail in~\cite{LRRRSSV2}.

 There remains the question of the additional SM singlet states that were
required to cancel the grav-grav-$\Z{N}^R$ graviton anomaly. Their charges are
such that $\sum_{i}(s_{i}-1)$ is odd and so there must be at least one state,
$S$, with even $\Z{4}^R$ charge. 
However, the GS mechanism requires the presence of a light axion, and the
axino contribution turns out to cancel the grav-grav-$\Z{4}^R$ anomaly. The
minimal realization of our $\Z4^R$ symmetry is therefore a setting in which
supersymmetry is broken by the dilaton type multiplet $S$ containing the
axino/dilatino; details will be given elsewhere~\cite{LRRRSSV2}.
It is remarkable that gravitational anomalies lead us to introduce this sector,
such that supersymmetry is broken ``outside the MSSM'', consistently with
phenomenological requirements.  That is, the missing spin-1/2 field is needed to
give mass to the gravitino. As discussed above, the VEV of the hidden sector
superpotential represents an order parameter for $\Z{4}^R$ breaking.

 We can now determine the phenomenology of the $\Z4^{R}$ model after
supersymmetry breaking. The residual $\Z2$ matter parity ensures that the
renormalisable terms are identical to those of the usual MSSM with no baryon- or
lepton-number violating terms. It also ensures that the supersymmetric partners
of SM states can only be pair produced and that the LSP is stable and a dark
matter candidate. The $\mu$-term is of the same order as the other visible
sector supersymmetry breaking terms and thus the model has completely solved the
$\mu$-problem. The lowest order baryon- and lepton-number violating terms occur
at dimension five, but these operators are strongly suppressed by a
non-perturbative factor of
$\mathcal{O}(\langle\mathscr{W}\rangle/M_\mathrm{P}^{4}$). For the case of SUGRA
this is of $\mathcal{O}(m_{3/2}/M_\mathrm{P}^2)$ and is negligible such that the
dimension six proton decay operators will be dominant.

Finally we should consider the cosmological implications of the model. Since it
involves a spontaneously broken discrete symmetry one must worry about domain
walls being produced in the early universe and dominating the energy density
today. A general discussion of walls resulting from the breaking of discrete
symmetries has recently appeared \cite{Dine:2010eb} (see
also~\cite{Abel:1995wk}) and we refer the reader to it for details appropriate
to various choices of the messenger scale. For the case of gravity mediation the
domain walls form at the intermediate scale of $\mathcal{O}(10^{12}\gev)$. 
Provided the Hubble scale during inflation is below this scale, domain walls
have sufficient time to form and then they will be inflated away. However to
avoid recreating them it is necessary that the reheat temperature after
inflation should be less than $\mathcal{O}(10^{12}\gev)$.  Given that the
gravitino and thermal moduli destabilization~\cite{Buchmuller:2004xr} bounds
require a reheat temperature much below this we conclude that the domain walls
in this case do not introduce a significant new problem for SUGRA.  Of course
one must still deal with the Polonyi problem \cite{Coughlan:1983ci} associated
with the energy released if there are light moduli fields but this is not
affected by having an underlying $\Z4^{R}$ symmetry. 

\section{String theory realization}
\label{sec:StringExample}

Compactified string theories often generate discrete gauge symmetries in the low
energy effective Lagrangian so it is appropriate to ask if they
can provide  the origin of the $\Z4^R$. In particular, (heterotic) orbifolds are
known to incorporate discrete $R$ symmetries in their effective field theory
description. These $R$ symmetries are discrete remnants of the Lorentz group of
compact space. Specifically, some of these constructions exhibit a $\Z4^R$,
reflecting the discrete rotational symmetry of a \Z2 orbifold plane
$\mathbbm{T}^2/\Z2$. 

Making extensive use of the methods to determine the remnant symmetries
described in \cite{Petersen:2009ip}, we were able to find examples realizing the
$\Z4^R$ just introduced, based on the $\Z2\times\Z2$ orbifold model derived in
\cite{Blaszczyk:2009in} and similar models, which have three $\mathbbm{T}^2/\Z2$
planes. These models have vacua with the exact MSSM spectrum, a large top Yukawa
coupling, a non-trivial hidden sector etc. In what follows, we briefly discuss a
vacuum exhibiting the $\Z4^R$ discussed above, defering a detailed description
to a subsequent publication \cite{LRRRSSV2}.

We found a configuration in the model \cite{Blaszczyk:2009in} in which the
$\Z4^R$ arises as a mix of the orbifold $\Z4^R$ symmetries and other symmetries.
The configuration is defined by assigning VEVs to some standard model singlet
fields, which break the symmetries at the orbifold point, \emph{i.e.}\ discrete
$R$ symmetries, discrete symmetries coming from the space group selection rules
and the gauged continuous symmetries are broken down to
$G_\mathrm{SM}\times\Z4^R\times\Z2$. The VEVs also provide mass terms for the
exotics, which are massless at the orbifold point, and allow us to cancel the
one-loop Fayet-Iliopoulos term associated with the one anomalous
$\U1_\mathrm{anom}$ of the heterotic orbifold model.  

Matter fields, of which we obtain precisely three generations, are identified as
fields with $\Z4^R$ charge 1.
There is one massless Higgs pair (with $R$ charge 0) at the perturbative level.
Unfortunately, the additional \Z2, which we cannot break, forbids some Yukawa
couplings such that the charged lepton and $d$-type Yukawa couplings $Y_e$ and
$Y_d$ have rank 2. 

The presence of the unwanted \Z2 shows that the vacuum is most likely not fully
realistic. Nevertheless our findings imply that the $\Z4^R$, which we have
identified solely by bottom-up considerations, can arise in potentially realistic
string compactifications, where the symmetry has a clear geometrical
interpretation. These models have an exact matter parity, a built-in solution to
the $\mu$-problem and do not suffer from the dimension five proton decay
problems. As they are string-derived (and hence UV complete), we can specify the
non-perturbative effects that appear to violate the `anomalous' $\Z4^R$ in more
detail. The (universal) anomalies are canceled by the Green-Schwarz mechanism.
That is, the imaginary part of the dilaton $S$ shifts under the discrete
transformations. As a consequence, terms of the form
\begin{align}\label{eq:WnpHiggs}
\mathscr{W}_\mathrm{np}&\supset~\mathrm{e}^{-8\pi^2\,S}\left(A\,H\,\Bar{H}
+\kappa_{ijk\ell}\,Q_i\,Q_j\,Q_k\,L_\ell\right)\;,
\end{align}
where $A$ and $\kappa_{ijk\ell}$ are constants built of some VEV fields, are
$\Z4^R$ covariant, \emph{i.e.}\ have $R$ charge 2. Such terms can be interpreted as
being a consequence of some hidden sector strong dynamics (the model
under consideration has a hidden \SU3). Assuming that the scale of supersymmetry
breakdown and the expectation value of the superpotential are related to this
strong dynamics, we obtain a $\mu$-term of the order of the gravitino mass
(cf.\ \cite{Brummer:2010fr}) and
coefficients of the dimension five proton decay operators as small as
$\sim10^{-15}/M_\mathrm{P}$, \emph{i.e.}\ well below experimental
bounds \cite{Hinchliffe:1992ad}.

\section{Summary}
\label{sec:Conclusions}

 Supersymmetric extensions of the SM promise to eliminate the hierarchy problem.
However they also introduce  serious potential problems and to be viable they
must evade the $\mu$-problem and the problem associated with new baryon- and
lepton-number violating terms. This suggests that there should be an additional
underlying symmetry capable of controlling these terms. 

 In this paper we have considered the anomaly free Abelian discrete symmetries
that forbid the $\mu$-term perturbatively. Remarkably, if one also requires that
the symmetry should simply be consistent with \SO{10} unification, there is a
unique solution, a $\Z{4}^R$ discrete $R$-symmetry. At perturbative order it
forbids dimension four and five baryon- and lepton-number violating terms. Being
anomalous in the absence of Green-Schwarz terms one may expect the symmetry to
be broken non-perturbatively, most likely through a gaugino condensate. At the
non-perturbative level, both the $\mu$-term and dimension five proton decay
operators may arise, while the dimension four operators are still forbidden by a
non-anomalous subgroup of $\Z{4}^R$ that is equivalent to matter parity. The
magnitude of the dimension five terms is such that the limits on nucleon decay
are readily satisfied. Inflation can ensure that the domain walls that are
produced when the discrete symmetry is broken are not significant provided a
mild upper bound on the inflation scale is satisfied. 

 Discrete $R$-symmetries can result from compactified string models as discrete
remnants of the Lorentz group of the compact space. We illustrated this in the
context of a semi-realistic orbifold model and showed how it gives rise to the
MSSM spectrum below the string scale with a $\Z{4}^R$ discrete $R$-symmetry.

\subsection*{Acknowledgments}

 This research was supported by the DFG cluster of excellence Origin and
Structure of the Universe, the \mbox{SFB-Transregio} 27 ``Neutrinos and
Beyond'', LMUExcellent and the Graduiertenkolleg ``Particle Physics at the
Energy Frontier of New Phenomena'' by Deutsche Forschungsgemeinschaft (DFG). 
S.R.\ acknowledges partial support from DOE grant DOE/ER/01545-887, and the CERN, IMPRS
and TUM visitor programs. The work was partially supported by the EU RTN grant
UNILHC 23792. H.M.L.\ is supported by the Korean-CERN fellowship.

\providecommand{\bysame}{\leavevmode\hbox to3em{\hrulefill}\thinspace}

\end{document}